\documentclass{article}%
\usepackage{graphicx}
\usepackage{amsmath}
\usepackage{amsfonts}
\usepackage{amssymb}%
\newtheorem{theorem}{Theorem}

\newenvironment{proof}[1][Proof]{\textbf{#1.} }{\ \rule{0.5em}{0.5em}}
\begin{document}

\title{Synthetic Vector Analysis III\\From Vector Analysis to Differential Forms}
\author{Hirokazu Nishimura\\Institute of Mathematics, University of Tsukuba\\Tsukuba, Ibaraki, 305-8571, Japan}
\maketitle

\begin{abstract}
In our previous paper [International Journal of Theoretical Physics,
\textbf{41} (2002), 1165-1190] we have shown, following the tradition of
synthetic differential geometry, that $\mathrm{div}$ and $\mathrm{rot}$ are
uniquely determined, so long as we require that the divergence theorem and the
Stokes theorem should hold on the infinitesimal level. In this paper we will
simplify the discussion considerably in terms of differential forms, leading
to the natural derivation of exterior differentiation in the usual form.

\end{abstract}

\section{Introduction}

Vector analysis presupposes dogmatically that every physical quantity is
either a scalar or a vector, excluding the possibility of tensors as natural
physical quantities. In vector analysis, the force and the flux are equally
vector fields, but, to tell the truth, the former is a field of tensors of
degree $1$, while the latter is a field of skew-symmetric tensors of degree
$2$. In electromagnetism, $\mathbf{E}$ and $\mathbf{B}$ are fields of tensors
of degree $1$, while $\mathbf{D}$ and $\mathbf{H}$ are fields of
skew-symmetric tensors of degree $2$. It is not desirable to apply
$\mathrm{div}$ to $\mathbf{E}$ or $\mathbf{B}$, though $\mathrm{curl}$ is
indeed applicable to both of them. It is not desirable to apply $\mathrm{curl}%
$ to $\mathbf{D}$ or $\mathbf{H}$, while $\mathrm{div}$ is indeed applicable
to both of them. Since $\mathbf{E}$ and $\mathbf{D}$ as well as $\mathbf{B}$
and $\mathbf{H}$ are proportional in the vacuum, the confusion is apt to occur
and develop ! Some physicists even insist wrongly that the CGS system of
units, in which $\varepsilon_{0}=\mu_{0}=1$ holds, is superior to the MKSA
system of units.

Nowadays the number of textbooks on elementary physics (elementary
electromagnetism in particular) using differential forms in place of vector
analysis is increasing, though there are still only a few. It is easy to give
a dictionary of vector analysis into the framework of differential forms, so
that vector analysis is really to be absorbed into the calculus of
differential forms. Nevertheless vector analysis is still popular among
physicists and students of physics, mainly because vector analysis is highly
intuitive, while the calculus of differential forms is not. The exterior
differentiation in the calculus of differential forms is usually given as a
decree without taking care of its intuitive or physical foundations at all.

What is easily forgotten, such geniuses as Newton and Leibniz discussed
advanced calculus in terms of nilpotent infinitesimals without using limits at
all. It was in the 19th century, in the midst of the industrial revolution,
that advanced calculus was reformulated in terms of limits, while nilpotent
infinitesimals were intentionally neglected as anathema. Synthetic
differential geometry, born in the middle of the 20th century, suceeded in
reviving nilpotent infinitesimals in advanced calculus and differential
geometry without hurting mathematical rigor at all. Newton and Leibniz saw
nilpotent infinitesimals not in this world but in another world, and the 20th
century witnessed powerful gadgets, such as seen in forcing techniques of set
theory, sheaf theory and topos theory, by which many other mathematically
meaningful worlds can be coherently constructed. Synthetic differential
geometry regained the natural meaning of exterior differentiation in
differential forms. The principal objective in this paper is to convince
physicists that the exterior differentiation is well motivated, just as
$\mathrm{div}$ and $\mathrm{curl}$ in vector analysis. The paper is more
expository than anything else. We have tried to help physicists understand how
naturally vector analysis develops into the calculus of differential forms.

\section{Preliminaries}

We assume that the reader is familiar with Chapter 1 of Lavendhomme \cite{l1}.
The set $\mathbb{R}$ of (extended) real numbers is required to abide by the
Kock-Lawvere axiom (cf. p.2 of \cite{l1}). We denote by $D$ the set of real
numbers whose squares vanish. The Kock-Lawvere axiom implies that, given a
mapping $\varphi:\mathbb{R}^{n}\rightarrow\mathbb{R}$ and $\mathbf{x}%
,\mathbf{a\in}\mathbb{R}^{n}$, there exists a unique $\varphi^{\prime
}(\mathbf{x})(\mathbf{a})\in\mathbb{R}^{n}$ such that
\[
\varphi(\mathbf{x}+\mathbf{a}d)-\varphi(\mathbf{x})=\varphi^{\prime
}(\mathbf{x})(\mathbf{a})d
\]
for any $d\in D$. It can be shown easily that the mapping $\mathbf{a\in
}\mathbb{R}^{n}\mapsto\varphi^{\prime}(\mathbf{x})(\mathbf{a})\in
\mathbb{R}^{n}$, which is to be regarded as the derivative of $\varphi$\ at
$\mathbf{x}$, is linear. The mapping $\varphi^{\prime}(\mathbf{x})$ goes as
follows:%
\[
\varphi^{\prime}(\mathbf{x})=\frac{\partial\varphi}{\partial x_{1}}%
(\mathbf{x})\mathbf{d}x_{1}+...+\frac{\partial\varphi}{\partial x_{n}%
}(\mathbf{x})\mathbf{d}x_{n}%
\]
We denote by $\mathbf{e}_{i}$%
\[
\left(
\begin{array}
[c]{c}%
0\\
\vdots\\
0\\
1\\
0\\
\vdots\\
0
\end{array}
\right)
\]
($1\leq i\leq n$), where $1$ is positioned at the $i$-th place. Given
$\gamma:D^{m+1}\rightarrow\mathbb{R}^{n}$, $e\in D$ and a natural number $i$
with $1\leq i\leq m+1$, we write $\gamma_{e}^{i}$ for the mapping
$(d_{1},...d_{m})\in D^{m}\mapsto\gamma(d_{1},...,d_{i-1},e,d_{i}%
,...,d_{m})\in\mathbb{R}^{n}$.

Let us consider the usual three-dimensional space $\mathbb{R}^{3}$, which is
the favorite space of vector analysis. Viewing the force $\mathbf{f}%
(\mathbf{x})$ at $\mathbf{x\in}\mathbb{R}^{3}$ as a vector in the usual way
should be called an \textit{idealistic} or \textit{Platonic} view of force.
Our \textit{pragmatic} or \textit{operational} view of force is to consider
how to measure $\mathbf{f}(\mathbf{x})$ experimentally. If we move from
$\mathbf{x}$\ to $\mathbf{x}+\mathbf{a}d$ infinitesimally with $\mathbf{a}%
\in\mathbb{R}^{3}$ and $d\in D$, we get the power $\mathbf{f}(\mathbf{x}%
)\cdot\mathbf{a}d$, where $\cdot$ denotes the inner product of vectors. Our
pragmatic view of force recommends that the force at $\mathbf{x}$\ should not
be $\mathbf{f}(\mathbf{x})$ but the linear mapping $\mathbf{a}\in
\mathbb{R}^{3}\mapsto\mathbf{f}(\mathbf{x})\cdot\mathbf{a\in}\mathbb{R}$. We
stress that $\mathbf{f}(\mathbf{x})$ is recognized via the linear mapping
$\mathbf{a}\in\mathbb{R}^{3}\mapsto\mathbf{f}(\mathbf{x})\cdot\mathbf{a\in
}\mathbb{R}$. This is our view of force as a tensor of degree $1$. Therefore a
field of forces is no other than a differential $1$-form from a mathematical viewpoint.

Let us consider a flow of air in $\mathbb{R}^{3}$, which is very often
represented by a field $\mathbf{f}$ of vectors. Our pragmatic view of flow
recommends that we should measure how much air passes in a unit time through
the infinitesimal parallelogram whose four vertices are $\mathbf{x}$,
$\mathbf{x}+\mathbf{a}d_{1}$, $\mathbf{x}+\mathbf{b}d_{2}$ and $\mathbf{x}%
+\mathbf{a}d_{1}+\mathbf{b}d_{2}$ with $\mathbf{x},\mathbf{a},\mathbf{b\in
}\mathbb{R}^{3}$ and $d_{1},d_{2}\in D$. The result is surely $\mathbf{f(x)}%
\cdot(\mathbf{a}\times\mathbf{b})d_{1}d_{2}$, where $\times$ stands for the
vector product. We would like to consider pragamatically that the
skew-symmetric bilinear mapping $(\mathbf{a},\mathbf{b})\in\mathbb{R}%
^{3}\times\mathbb{R}^{3}\mapsto\mathbf{f(x)}\cdot(\mathbf{a}\times
\mathbf{b})\in\mathbb{R}$ is no other than the mathematical representation of
the flow at $\mathbf{x}$. In this sense, the flow is represented by a field of
skew-symmetric tensors of degree $2$, namely, by a differential $2$-form.

We know well that every linear mapping from $\mathbb{R}^{3}$ to $\mathbb{R}$
is of the form $\alpha_{1}\mathbf{d}x+\alpha_{2}\mathbf{d}y+\alpha
_{3}\mathbf{d}z$ with $\alpha_{1},\alpha_{2},\alpha_{3}\in\mathbb{R}$, while
every skew-symmetric bilinear mapping from $\mathbb{R}^{3}\times\mathbb{R}%
^{3}$ to $\mathbb{R}$ is of the form $\alpha_{1}\mathbf{d}y\wedge
\mathbf{d}z+\alpha_{2}\mathbf{d}z\wedge\mathbf{d}x+\alpha_{3}\mathbf{d}%
x\wedge\mathbf{d}y$. We know well that every skew-symmetric trilinear mapping
from $\mathbb{R}^{3}\times\mathbb{R}^{3}\times\mathbb{R}^{3}$ to $\mathbb{R}$
is of the form $\alpha\mathbf{d}x\wedge\mathbf{d}y\wedge\mathbf{d}z$ with
$\alpha\in\mathbb{R}$. More generally, every skew-symmetric $k$-linear mapping
from $\underset{k}{\underbrace{\mathbb{R}^{n}\times\mathbb{\cdot\cdot\cdot
}\times\mathbb{R}^{n}}}$ to $\mathbb{R}$ is of the form%
\[
\sum_{1\leq i_{1}<...<i_{k}\leq n}\alpha_{i_{1},...,i_{k+1}}\mathbf{d}%
x_{i_{1}}\wedge...\wedge\mathbf{d}x_{i_{k}}%
\]

In vector analysis, the operators $\mathrm{div}$ and $\mathrm{rot}$ are
determined uniquely so that the divergence theorem and the Stokes' theorem
should hold on the infinitesimal level respectively. In the same way, the
exterior differentiation from a differential $k$-form to a differential
$(k+1)$-form is determined uniquely so that Stokes' theorem should hold on the
infinitesimal level. The principal objective in this paper is to give a lucid
explanation on these facts as elementarily as possible from the standpoint of
synthetic differential geometry, while avoiding the utmost generality, which
would  usually be liable to defy ordinary physicists.

\section{The Fundamental Theorem for Gradient}

\begin{theorem}
Let $\varphi$ be a scalar field on $\mathbb{R}^{3}$. Let $t:d\in
D\mapsto\mathbf{x}+\mathbf{a}d$ be a tangent vector at $\mathbf{x}$ on
$\mathbb{R}^{3}$. Let $e\in D$. Then we have
\[
\int_{\partial(t;e)}\varphi=\int_{(t;e)}\mathbf{d}\varphi
\]
where
\[
\partial(t;e)=(\mathbf{x}+\mathbf{a}e)-(\mathbf{x})
\]

\end{theorem}

\begin{proof}
This is no other than the definition of $\mathbf{d}\varphi=\varphi^{\prime}$,
namely,
\[
\varphi(\mathbf{x}+\mathbf{a}e)-\varphi(\mathbf{x})=\varphi^{\prime
}(\mathbf{x})(\mathbf{a})e
\]

\end{proof}

\section{The Fundamental Theorem for Rotation}

\begin{theorem}
Let $\omega=f\mathbf{d}x+g\mathbf{d}y+h\mathbf{d}z$ be a differential $1$-form
on $\mathbb{R}^{3}$. Let $\gamma:(d_{1},d_{2})\in D^{2}\mapsto\mathbf{x}%
+\mathbf{a}d_{1}+\mathbf{b}d_{2}$ be an infinitesimal parallelogram at
$\mathbf{x}$ on $\mathbb{R}^{3}$. Let $(e_{1},e_{2})\in D^{2}$. Then we have
\[
\int_{\partial(\gamma;e_{1},e_{2})}\omega=\int_{(\gamma;e_{1},e_{2}%
)}\mathbf{d}\omega
\]
where
\begin{align*}
&  \partial(\gamma;e_{1},e_{2})\\
&  =(\gamma_{0}^{2};e_{1})+(\gamma_{e_{1}}^{1};e_{2})-(\gamma_{e_{2}}%
^{2};e_{1})-(\gamma_{0}^{1};e_{2})
\end{align*}
and
\[
\mathbf{d}\omega=(\frac{\partial h}{\partial y}-\frac{\partial g}{\partial
z})\mathbf{d}y\wedge\mathbf{d}z+(\frac{\partial f}{\partial z}-\frac{\partial
h}{\partial x})\mathbf{d}z\wedge\mathbf{d}x+(\frac{\partial g}{\partial
x}-\frac{\partial f}{\partial y})\mathbf{d}x\wedge\mathbf{d}y
\]

\end{theorem}

\begin{proof}
We have
\begin{align*}
&  \int_{\partial(\gamma;e_{1},e_{2})}\omega\\
&  =\int_{(\gamma_{0}^{2};e_{1})}\omega+\int_{(\gamma_{e_{1}}^{1};e_{2}%
)}\omega-\int_{(\gamma_{e_{2}}^{2};e_{1})}\omega-\int_{(\gamma_{0}^{1};e_{2}%
)}\omega\\
&  =\left\{  f(\mathbf{x})a_{1}+g(\mathbf{x})a_{2}+h(\mathbf{x})a_{3}\right\}
e_{1}\\
&  +\left\{  f(\mathbf{x+a}e_{1})b_{1}+g(\mathbf{x+a}e_{1})b_{2}%
+h(\mathbf{x+a}e_{1})b_{3}\right\}  e_{2}\\
&  -\left\{  f(\mathbf{x+b}e_{2})a_{1}+g(\mathbf{x+b}e_{2})a_{2}%
+h(\mathbf{x+b}e_{2})a_{3}\right\}  e_{1}\\
&  -\left\{  f(\mathbf{x})b_{1}+g(\mathbf{x})b_{2}+h(\mathbf{x})b_{3}\right\}
e_{2}\\
&  =\left\{  f^{\prime}(\mathbf{x})(\mathbf{a})b_{1}+g^{\prime}(\mathbf{x}%
)(\mathbf{a})b_{2}+h^{\prime}(\mathbf{x})(\mathbf{a})b_{3}\right\}  e_{1}%
e_{2}\\
&  -\left\{  f^{\prime}(\mathbf{x})(\mathbf{b})a_{1}+g^{\prime}(\mathbf{x}%
)(\mathbf{b})a_{2}+h^{\prime}(\mathbf{x})(\mathbf{b})a_{3}\right\}  e_{1}%
e_{2}\\
&  \left[
\begin{array}
[c]{c}%
\text{The first term delineated by }\left\{  {}\right\}  \text{ and followed
by }e_{1}e_{2}\text{\ is obtained by }\\
\text{combining the second term and the fourth of the preceeding formula, }\\
\text{while the second term delineated by }\left\{  {}\right\}  \text{ and
followed by }e_{1}e_{2}\text{\ is obtained }\\
\text{by combining the first term and the third of the preceeding formula.}%
\end{array}
\right] \\
&  =\left\{  \left(
\begin{array}
[c]{c}%
f^{\prime}(\mathbf{x})(\mathbf{a})\\
g^{\prime}(\mathbf{x})(\mathbf{a})\\
h^{\prime}(\mathbf{x})(\mathbf{a})
\end{array}
\right)  \cdot\mathbf{b-}\left(
\begin{array}
[c]{c}%
f^{\prime}(\mathbf{x})(\mathbf{b})\\
g^{\prime}(\mathbf{x})(\mathbf{b})\\
h^{\prime}(\mathbf{x})(\mathbf{b})
\end{array}
\right)  \cdot\mathbf{a}\right\}  e_{1}e_{2}%
\end{align*}
Let $\varphi:\mathbb{R}^{3}\times\mathbb{R}^{3}\rightarrow\mathbb{R}$ be the
mapping
\begin{align*}
&  \varphi(\mathbf{a},\mathbf{b})\\
&  =\left(
\begin{array}
[c]{c}%
f^{\prime}(\mathbf{x})(\mathbf{a})\\
g^{\prime}(\mathbf{x})(\mathbf{a})\\
h^{\prime}(\mathbf{x})(\mathbf{a})
\end{array}
\right)  \cdot\mathbf{b-}\left(
\begin{array}
[c]{c}%
f^{\prime}(\mathbf{x})(\mathbf{b})\\
g^{\prime}(\mathbf{x})(\mathbf{b})\\
h^{\prime}(\mathbf{x})(\mathbf{b})
\end{array}
\right)  \cdot\mathbf{a}%
\end{align*}
for any $(\mathbf{a},\mathbf{b})\in\mathbb{R}^{3}\times\mathbb{R}^{3}$, so
that
\[
\int_{\partial(\gamma;e_{1},e_{2})}\omega=\varphi(\mathbf{a},\mathbf{b}%
)e_{1}e_{2}%
\]
Then it is easy to see that $\varphi$ is a skew-symmetric bilinear mapping, so
that $\varphi$ is of the form
\[
\varphi=\alpha_{1}\mathbf{d}y\wedge\mathbf{d}z+\alpha_{2}\mathbf{d}%
z\wedge\mathbf{d}x+\alpha_{3}\mathbf{d}x\wedge\mathbf{d}y
\]
with $\alpha_{i}\in\mathbb{R}$ ($i=1,2,3$). By taking

\begin{enumerate}
\item $\mathbf{a}=\mathbf{e}_{2}$ and $\mathbf{b}=\mathbf{e}_{3}$

\item $\mathbf{a}=\mathbf{e}_{3}$ and $\mathbf{b}=\mathbf{e}_{1}$, or

\item $\mathbf{a}=\mathbf{e}_{1}$ and $\mathbf{b}=\mathbf{e}_{2}$,
\end{enumerate}

we get
\begin{align*}
\alpha_{1}  &  =\frac{\partial h}{\partial y}(\mathbf{x})-\frac{\partial
g}{\partial z}(\mathbf{x})\\
\alpha_{2}  &  =\frac{\partial f}{\partial z}(\mathbf{x})-\frac{\partial
h}{\partial x}(\mathbf{x})\\
\alpha_{3}  &  =\frac{\partial g}{\partial x}(\mathbf{x})-\frac{\partial
f}{\partial y}(\mathbf{x})
\end{align*}
easily. This completes the proof.
\end{proof}

\section{The Fundamental Theorem for Divergence}

\begin{theorem}
Let $\omega=f\mathbf{d}y\wedge\mathbf{d}z\wedge+g\mathbf{d}z\wedge
\mathbf{d}x+h\mathbf{d}x\wedge\mathbf{d}y$ be a differential $2$-form on
$\mathbb{R}^{3}$. Let $\gamma:(d_{1},d_{2},d_{3})\in D^{3}\mapsto
\mathbf{x}+\mathbf{a}d_{1}+\mathbf{b}d_{2}+\mathbf{c}d_{3}$ be an
infinitesimal parallelepiped at $\mathbf{x}$ on $\mathbb{R}^{3}$. Let
$(e_{1},e_{2},e_{3})\in D^{3}$. Then we have
\[
\int_{\partial(\gamma;e_{1},e_{2},e_{3})}\omega=\int_{(\gamma;e_{1}%
,e_{2},e_{3})}\mathbf{d}\omega
\]
where
\begin{align*}
&  \partial(\gamma;e_{1},e_{2},e_{3})\\
&  =-(\gamma_{0}^{1};e_{2},e_{3})+(\gamma_{e_{1}}^{1};e_{2},e_{3})+(\gamma
_{0}^{2};e_{1},e_{3})-(\gamma_{e_{2}}^{2};e_{1},e_{3})\\
&  -(\gamma_{0}^{3};e_{1},e_{2})+(\gamma_{e_{3}}^{3};e_{1},e_{2})
\end{align*}
and
\[
\mathbf{d}\omega=(\frac{\partial f}{\partial x}-\frac{\partial g}{\partial
y}+\frac{\partial h}{\partial z})\mathbf{d}x\wedge\mathbf{d}y\wedge
\mathbf{d}z
\]

\end{theorem}

\begin{proof}
We have
\begin{align*}
&  \int_{\partial(\gamma;e_{1},e_{2},e_{3})}\omega\\
&  =-\int_{(\gamma_{0}^{1};e_{2},e_{3})}\omega+\int_{(\gamma_{e_{1}}^{1}%
;e_{2},e_{3})}\omega+\int_{(\gamma_{0}^{2};e_{1},e_{3})}\omega-\int
_{(\gamma_{e_{2}}^{2};e_{1},e_{3})}\omega-\int_{(\gamma_{0}^{3};e_{1},e_{2}%
)}\omega+\int_{(\gamma_{e_{3}}^{3};e_{1},e_{2})}\omega\\
&  =-\left\{  f(\mathbf{x})\left|
\begin{array}
[c]{cc}%
b_{2} & c_{2}\\
b_{3} & c_{3}%
\end{array}
\right|  +g(\mathbf{x})\left|
\begin{array}
[c]{cc}%
b_{3} & c_{3}\\
b_{1} & c_{1}%
\end{array}
\right|  +h(\mathbf{x})\left|
\begin{array}
[c]{cc}%
b_{1} & c_{1}\\
b_{2} & c_{2}%
\end{array}
\right|  \right\}  e_{2}e_{3}\\
&  +\left\{  f(\mathbf{x+a}e_{1})\left|
\begin{array}
[c]{cc}%
b_{2} & c_{2}\\
b_{3} & c_{3}%
\end{array}
\right|  +g(\mathbf{x+a}e_{1})\left|
\begin{array}
[c]{cc}%
b_{3} & c_{3}\\
b_{1} & c_{1}%
\end{array}
\right|  +h(\mathbf{x+a}e_{1})\left|
\begin{array}
[c]{cc}%
b_{1} & c_{1}\\
b_{2} & c_{2}%
\end{array}
\right|  \right\}  e_{2}e_{3}\\
&  +\left\{  f(\mathbf{x})\left|
\begin{array}
[c]{cc}%
a_{2} & c_{2}\\
a_{3} & c_{3}%
\end{array}
\right|  +g(\mathbf{x})\left|
\begin{array}
[c]{cc}%
a_{3} & c_{3}\\
a_{1} & c_{1}%
\end{array}
\right|  +h(\mathbf{x})\left|
\begin{array}
[c]{cc}%
a_{1} & c_{1}\\
a_{2} & c_{2}%
\end{array}
\right|  \right\}  e_{1}e_{3}\\
&  -\left\{  f(\mathbf{x+b}e_{1})\left|
\begin{array}
[c]{cc}%
a_{2} & c_{2}\\
a_{3} & c_{3}%
\end{array}
\right|  +g(\mathbf{x+b}e_{1})\left|
\begin{array}
[c]{cc}%
a_{3} & c_{3}\\
a_{1} & c_{1}%
\end{array}
\right|  +h(\mathbf{x+b}e_{1})\left|
\begin{array}
[c]{cc}%
a_{1} & c_{1}\\
a_{2} & c_{2}%
\end{array}
\right|  \right\}  e_{1}e_{3}\\
&  -\left\{  f(\mathbf{x})\left|
\begin{array}
[c]{cc}%
a_{2} & b_{2}\\
a_{3} & b_{3}%
\end{array}
\right|  +g(\mathbf{x})\left|
\begin{array}
[c]{cc}%
a_{3} & b_{3}\\
a_{1} & b_{1}%
\end{array}
\right|  +h(\mathbf{x})\left|
\begin{array}
[c]{cc}%
a_{1} & b_{1}\\
a_{2} & b_{2}%
\end{array}
\right|  \right\}  e_{1}e_{2}\\
&  +\left\{  f(\mathbf{x+c}e_{3})\left|
\begin{array}
[c]{cc}%
a_{2} & b_{2}\\
a_{3} & b_{3}%
\end{array}
\right|  +g(\mathbf{x+c}e_{3})\left|
\begin{array}
[c]{cc}%
a_{3} & b_{3}\\
a_{1} & b_{1}%
\end{array}
\right|  +h(\mathbf{x+c}e_{3})\left|
\begin{array}
[c]{cc}%
a_{1} & b_{1}\\
a_{2} & b_{2}%
\end{array}
\right|  \right\}  e_{1}e_{2}\\
&  =\left\{  f^{\prime}(\mathbf{x})(\mathbf{a})\left|
\begin{array}
[c]{cc}%
b_{2} & c_{2}\\
b_{3} & c_{3}%
\end{array}
\right|  +g^{\prime}(\mathbf{x})(\mathbf{a})\left|
\begin{array}
[c]{cc}%
b_{3} & c_{3}\\
b_{1} & c_{1}%
\end{array}
\right|  +h^{\prime}(\mathbf{x})(\mathbf{a})\left|
\begin{array}
[c]{cc}%
b_{1} & c_{1}\\
b_{2} & c_{2}%
\end{array}
\right|  \right\}  e_{1}e_{2}e_{3}\\
&  -\left\{  f^{\prime}(\mathbf{x})(\mathbf{b})\left|
\begin{array}
[c]{cc}%
a_{2} & c_{2}\\
a_{3} & c_{3}%
\end{array}
\right|  +g^{\prime}(\mathbf{x})(\mathbf{b})\left|
\begin{array}
[c]{cc}%
a_{3} & c_{3}\\
a_{1} & c_{1}%
\end{array}
\right|  +h^{\prime}(\mathbf{x})(\mathbf{b})\left|
\begin{array}
[c]{cc}%
a_{1} & c_{1}\\
a_{2} & c_{2}%
\end{array}
\right|  \right\}  e_{1}e_{2}e_{3}\\
&  +\left\{  f^{\prime}(\mathbf{x})(\mathbf{c})\left|
\begin{array}
[c]{cc}%
b_{2} & c_{2}\\
b_{3} & c_{3}%
\end{array}
\right|  +g^{\prime}(\mathbf{x})(\mathbf{c})\left|
\begin{array}
[c]{cc}%
b_{3} & c_{3}\\
b_{1} & c_{1}%
\end{array}
\right|  +h^{\prime}(\mathbf{x})(\mathbf{c})\left|
\begin{array}
[c]{cc}%
b_{1} & c_{1}\\
b_{2} & c_{2}%
\end{array}
\right|  \right\}  e_{1}e_{2}e_{3}\\
&  =\left\{  \left|
\begin{array}
[c]{ccc}%
\begin{array}
[c]{c}%
f^{\prime}(\mathbf{x})(\mathbf{a})\\
g^{\prime}(\mathbf{x})(\mathbf{a})\\
h^{\prime}(\mathbf{x})(\mathbf{a})
\end{array}
& \mathbf{b} & \mathbf{c}%
\end{array}
\right|  +\left|
\begin{array}
[c]{ccc}%
\mathbf{a} &
\begin{array}
[c]{c}%
f^{\prime}(\mathbf{x})(\mathbf{b})\\
g^{\prime}(\mathbf{x})(\mathbf{b})\\
h^{\prime}(\mathbf{x})(\mathbf{b})
\end{array}
& \mathbf{c}%
\end{array}
\right|  +\left|
\begin{array}
[c]{ccc}%
\mathbf{a} & \mathbf{b} &
\begin{array}
[c]{c}%
f^{\prime}(\mathbf{x})(\mathbf{c})\\
g^{\prime}(\mathbf{x})(\mathbf{c})\\
h^{\prime}(\mathbf{x})(\mathbf{c})
\end{array}
\end{array}
\right|  \right\}  e_{1}e_{2}e_{3}%
\end{align*}
Let $\varphi:\mathbb{R}^{3}\times\mathbb{R}^{3}\times\mathbb{R}^{3}%
\rightarrow\mathbb{R}$ be the mapping
\begin{align*}
&  \varphi(\mathbf{a},\mathbf{b},\mathbf{c})\\
&  =\left|
\begin{array}
[c]{ccc}%
\begin{array}
[c]{c}%
f^{\prime}(\mathbf{x})(\mathbf{a})\\
g^{\prime}(\mathbf{x})(\mathbf{a})\\
h^{\prime}(\mathbf{x})(\mathbf{a})
\end{array}
& \mathbf{b} & \mathbf{c}%
\end{array}
\right|  +\left|
\begin{array}
[c]{ccc}%
\mathbf{a} &
\begin{array}
[c]{c}%
f^{\prime}(\mathbf{x})(\mathbf{b})\\
g^{\prime}(\mathbf{x})(\mathbf{b})\\
h^{\prime}(\mathbf{x})(\mathbf{b})
\end{array}
& \mathbf{c}%
\end{array}
\right|  +\left|
\begin{array}
[c]{ccc}%
\mathbf{a} & \mathbf{b} &
\begin{array}
[c]{c}%
f^{\prime}(\mathbf{x})(\mathbf{c})\\
g^{\prime}(\mathbf{x})(\mathbf{c})\\
h^{\prime}(\mathbf{x})(\mathbf{c})
\end{array}
\end{array}
\right|
\end{align*}
for any $(\mathbf{a},\mathbf{b},\mathbf{c})\in\mathbb{R}^{3}\times
\mathbb{R}^{3}\times\mathbb{R}^{3}$, so that
\[
\int_{\partial(\gamma;e_{1},e_{2},e_{3})}\omega=\varphi(\mathbf{a}%
,\mathbf{b},\mathbf{c})e_{1}e_{2}e_{3}%
\]
Then it is easy to see that $\varphi$ is a skew-symmetric trilinear mapping,
so that $\varphi$ is of the form
\[
\varphi=\alpha\mathbf{d}x\wedge\mathbf{d}y\wedge\mathbf{d}z
\]
with $\alpha\in\mathbb{R}$. By taking $\mathbf{a}=\mathbf{e}_{1}$,
$\mathbf{b}=\mathbf{e}_{2}$ and $\mathbf{c}=\mathbf{e}_{3}$, we get
\[
\alpha=\frac{\partial f}{\partial x}(\mathbf{x})+\frac{\partial g}{\partial
y}(\mathbf{x})+\frac{\partial h}{\partial z}(\mathbf{x})
\]
easily. This completes the proof.
\end{proof}

\section{The Fundamental Theorem for Exterior Differentiation}

\begin{theorem}
Let $\omega=\sum_{1\leq i_{1}<...<i_{k}\leq n}f_{i_{1},...,i_{k}}%
\mathbf{d}x_{i_{1}}\wedge...\wedge\mathbf{d}x_{i_{k}}$ be a differential
$k$-form on $\mathbb{R}^{n}$. Let $\gamma:(d_{1},...,d_{k+1})\in
D^{k+1}\mapsto\mathbf{a}^{1}d_{1}+\mathbf{...}+\mathbf{a}^{k+1}d_{k+1}$ be an
infinitesimal $(k+1)$-parallelepiped at $\mathbf{x}$ on $\mathbb{R}^{n}$. Let
$(e_{1},...,e_{k+1})\in D^{k+1}$. Then we have
\[
\int_{\partial(\gamma;e_{1},...,e_{k+1})}\omega=\int_{(\gamma;e_{1}%
,...,e_{k+1})}\mathbf{d}\omega
\]
where
\begin{align*}
&  \partial(\gamma;e_{1},...,e_{k+1})\\
&  =\sum_{i=1}^{k+1}(-1)^{i}\left\{  (\gamma_{0}^{i};e_{1},...,\widehat{e_{i}%
},...,e_{k+1})-(\gamma_{e_{i}}^{i};e_{1},...,\widehat{e_{i}},...,e_{k+1}%
)\right\}
\end{align*}
and
\[
\mathbf{d}\omega=\sum_{1\leq i_{1}<...<i_{k+1}\leq n}\left(  \sum_{j=1}%
^{k+1}(-1)^{j+1}\frac{\partial f_{i_{1},...,\widehat{,i_{j}},...,i_{k+1}}%
}{\partial x_{i_{j}}}\right)  \mathbf{d}x_{i_{1}}\wedge...\wedge
\mathbf{d}x_{i_{k+1}}%
\]

\end{theorem}

\begin{proof}
We have
\begin{align*}
&  \int_{\partial(\gamma;e_{1},...,e_{k+1})}\omega\\
&  =\sum_{i=1}^{k+1}(-1)^{i}\left\{  \int_{(\gamma_{0}^{i};e_{1}%
,...,\widehat{e_{i}},...,e_{k+1})}\omega-\int_{(\gamma_{e_{i}}^{i}%
;e_{1},...,\widehat{e_{i}},...,e_{k+1})}\omega\right\} \\
&  =\sum_{1\leq i_{1}<...<i_{k}\leq n}\sum_{i=1}^{k+1}(-1)^{i}\left\{
\begin{array}
[c]{c}%
f_{i_{1},...,i_{k}}(\mathbf{x})\left\vert
\begin{array}
[c]{cccccc}%
a_{i_{1}}^{1} & \cdots & a_{i_{1}}^{i-1} & a_{i_{1}}^{i+1} & \cdots &
a_{i_{1}}^{k+1}\\
\vdots & \vdots & \vdots & \vdots & \vdots & \vdots\\
a_{i_{k}}^{1} & \cdots & a_{i_{k}}^{i-1} & a_{i_{k}}^{i+1} & \cdots &
a_{i_{k}}^{k+1}%
\end{array}
\right\vert e_{1}...\widehat{e_{i}}...e_{k+1}-\\
f_{i_{1},...,i_{k}}(\mathbf{x+a}^{i}e_{i})\left\vert
\begin{array}
[c]{cccccc}%
a_{i_{1}}^{1} & \cdots & a_{i_{1}}^{i-1} & a_{i_{1}}^{i+1} & \cdots &
a_{i_{1}}^{k+1}\\
\vdots & \vdots & \vdots & \vdots & \vdots & \vdots\\
a_{i_{k}}^{1} & \cdots & a_{i_{k}}^{i-1} & a_{i_{k}}^{i+1} & \cdots &
a_{i_{k}}^{k+1}%
\end{array}
\right\vert e_{1}...\widehat{e_{i}}...e_{k+1}%
\end{array}
\right\} \\
&  =\sum_{1\leq i_{1}<...<i_{k}\leq n}\sum_{i=1}^{k+1}(-1)^{i+1}%
f_{i_{1},...,i_{k}}^{\prime}(\mathbf{x})(\mathbf{a}^{i})\left\vert
\begin{array}
[c]{cccccc}%
a_{i_{1}}^{1} & \cdots & a_{i_{1}}^{i-1} & a_{i_{1}}^{i+1} & \cdots &
a_{i_{1}}^{k+1}\\
\vdots & \vdots & \vdots & \vdots & \vdots & \vdots\\
a_{i_{k}}^{1} & \cdots & a_{i_{k}}^{i-1} & a_{i_{k}}^{i+1} & \cdots &
a_{i_{k}}^{k+1}%
\end{array}
\right\vert e_{1}...e_{k+1}\\
&  =\sum_{1\leq i_{1}<...<i_{k}\leq n}\left\vert
\begin{array}
[c]{ccc}%
f_{i_{1},...,i_{k}}^{\prime}(\mathbf{x})(\mathbf{a}^{1}) & \cdots &
f_{i_{1},...,i_{k}}^{\prime}(\mathbf{x})(\mathbf{a}^{k+1})\\
a_{i_{1}}^{1} & \cdots & a_{i_{1}}^{k+1}\\
\vdots & \cdots & \vdots\\
a_{i_{k}}^{1} & \cdots & a_{i_{k}}^{k+1}%
\end{array}
\right\vert e_{1}...e_{k+1}%
\end{align*}
Let $\varphi:\underset{k+1}{\underbrace{\mathbb{R}^{n}\times...\times
\mathbb{R}^{n}}}\rightarrow\mathbb{R}$ be the mapping
\[
\varphi(\mathbf{a}^{1},...,\mathbf{a}^{k+1})=\sum_{1\leq i_{1}<...<i_{k}\leq
n}\left\vert
\begin{array}
[c]{ccc}%
f_{i_{1},...,i_{k}}^{\prime}(\mathbf{x})(\mathbf{a}^{1}) & \cdots &
f_{i_{1},...,i_{k}}^{\prime}(\mathbf{x})(\mathbf{a}^{k+1})\\
a_{i_{1}}^{1} & \cdots & a_{i_{1}}^{k+1}\\
\vdots & \cdots & \vdots\\
a_{i_{k}}^{1} & \cdots & a_{i_{k}}^{k+1}%
\end{array}
\right\vert
\]
for any $(\mathbf{a}^{1},...,\mathbf{a}^{k+1})\in\underset{k+1}{\underbrace
{\mathbb{R}^{n}\times...\times\mathbb{R}^{n}}}$, so that
\[
\int_{\partial(\gamma;e_{1},...,e_{k+1})}\omega=\varphi(\mathbf{a}%
^{1},...,\mathbf{a}^{k+1})e_{1}...e_{k+1}%
\]
Then it is easy to see that $\varphi$ is a skew-symmetric $(k+1)$-linear
mapping, so that $\varphi$ is of the form
\[
\varphi=\sum_{1\leq i_{1}<...<i_{k+1}\leq n}\alpha_{i_{1},...,i_{k+1}%
}\mathbf{d}x_{i_{1}}\wedge...\wedge\mathbf{d}x_{i_{k+1}}%
\]
with $\alpha_{i_{1},...,i_{k+1}}\in\mathbb{R}$ ($1\leq i_{1}<...<i_{k+1}\leq
n$). By taking $\mathbf{a}^{1}=\mathbf{e}_{i_{1}},\mathbf{a}^{2}%
=\mathbf{e}_{i_{2}},...,\mathbf{a}^{k+1}=\mathbf{e}_{i_{k+1}}$, we get
\[
\alpha_{i_{1},...,i_{k+1}}=\sum_{j=1}^{k+1}(-1)^{j+1}\frac{\partial
f_{i_{1},...,\widehat{,i_{j}},...,i_{k+1}}}{\partial x_{i_{j}}}%
\]
easily. This completes the proof.
\end{proof}


\begin{thebibliography}{9}                                                                                                %


\bibitem {b}Bell, John L.:A Primer of Infinitesimal Analysis, Cambridge
University Press, Cambridge, 1998.

\bibitem {ho}Hehl, Frierich W. and Obukhov, Yuri N.:Foundations of Classical
Electrodynamics, Birkh\"{a}user, Boston, 2003.

\bibitem {holm1}Holm, Darryl D.:Geometric Mechanics, 2 Vols, Imperial College
Press, London, 2008.

\bibitem {ki1}Kitano, M.:Maxwell Equations , Science, Tokyo, 2005 [in Japanese].

\bibitem {l1}Lavendhomme, R.:Basic Concepts of Synthetic Differential
Geometry, Kluwer, Dordrecht, 1996.

\bibitem {n1}Nishimura, H.:Synthetic vector analysis, International Journal of
Theoretical Physics, \textbf{41} (2002), 1165-1190.

\bibitem {n2}Nishimura, H.:Synthetic vector analysis II, International Journal
of Theoretical Physics, \textbf{43} (2004), 505-517.

\bibitem {o}Oliva, Waldyr Muniz:Geometric Mechanics, Springer, Berlin and
Heidelberg, 2002.
\end{thebibliography}
\end{document}